\documentclass[12pt]{article}
\pdfoutput=1

\usepackage{subfigure}  
\usepackage{graphicx}
\usepackage{amsfonts,amssymb}
\usepackage{epsfig}

\newcommand{\Dslash}{{\slash\!\!\!\!D}}

\newcommand{\sdv}[1]{\vec{\sigma}\cdot\vec{#1}}

\newcommand{\sh}{\sinh(2\lambda r)}

\title{\bf The 't~Hooft-Polyakov Monopole in the Presence of a 't~Hooft Operator}
\author{Sergey A. Cherkis\thanks{E-mail: cherkis@maths.tcd.ie}\\
\it School of Mathematics and Hamilton Mathematics Institute, \\
\it Trinity College, Dublin, Ireland
\rm
\and
Brian Durcan\thanks{E-mail: bdurcan@maths.tcd.ie}\\
\it School of Mathematics, Trinity College, Dublin, Ireland
}
\date{ }

\begin{document}
\begin{titlepage}

\renewcommand{\thepage}{ }

\maketitle

\begin{abstract}
We present explicit BPS field configurations representing one nonabelian monopole with one minimal weight 't~Hooft operator insertion.  We explore the $SO(3)$ and $SU(2)$  gauge groups.  

In the case of $SU(2)$ gauge group the minimal 't~Hooft operator can be completely screened by the monopole. If the gauge group is $SO(3),$ however,  such screening is impossible.  In the latter case we observe a  different effect of the gauge symmetry enhancement in the vicinity of the 't~Hooft operator.
\end{abstract}

\vspace{-6.5in}

\parbox{\linewidth}
{\small\hfill \shortstack{TCDMATH 07-22\\ \hfill HMI 07-09}}

\end{titlepage}

\section{Introduction}
't~Hooft operators \cite{'t Hooft:1977hy} play a central role in recent studies of the Montonen-Olive duality \cite{Montonen:1977sn} as electric-magetic duals of the Polyakov-Wilson operators \cite{Kapustin:2005py}.
Their significance as Hecke operators in the geometric Langlands program is elucidated in \cite{Kapustin:2006pk}.  

By a  't~Hooft operator we understand a line operator such that in the three-dimensional space transverse to the line it amounts to an insertion of a Dirac monopole imbedded into the gauge group in question.  In other words, in the vicinity of an insertion point in the three-dimensional space, we impose the following boundary conditions  \cite{Goddard:1976qe} on the gauge fields 
\begin{equation}\label{Eq:minimal}
F=\frac{B}{2}d\Omega_2,
\end{equation} 
where $d\Omega_2$ is the volume form of a unit sphere surrounding the point  and $B$ is the Lie algebra element satisfying 
$\exp(2\pi i B)={\mathbb I}.$
The 't~Hooft charge of such an operator takes values in $H^2(S^2,\pi_1(G)).$ It vanishes if the gauge group $G$ is $SU(2)$  and is ${\mathbb Z}/2{\mathbb  Z}$ valued if $G$ is $SO(3).$  Strictly speaking only the operators with nonzero 't~Hooft charge are significant in \cite{'t Hooft:1977hy}, but here we forgo this restriction and adopt the more general definition of \cite{Kapustin:2005py} and \cite{Kapustin:2006pk}. A minimal 't~Hooft operator is an insertion of a Dirac monopole of the lowest possible charge. Here we focus on such minimal operator insertions.

The exact analytic solutions we present in Section \ref{Sec:Solutions} probe minimal 't~Hooft operators with a monopole.  These solutions are new and can be interpreted as a nonlinear superposition of a Dirac monopole imbedded in the gauge group and a 't~Hooft-Polyakov monopole.  These configurations arise in various theories.  The most relevant to the electric-magnetic duality setting mentioned above is to view these as solutions in the maximally supersymmetric Yang-Mills theory in four dimensional space-time. 
In that case the Dirac monopole singularity is interpreted as a 't~Hooft time-like line operator. This is a codimension two operator.  Under the electric-magnetic duality such operators are mapped to Wilson line operators.  Since Wilson lines can terminate on quarks, a natural question posed by the duality is what  the dual of this phenomenon is.  What is the point, i.e. codimension three, operator that the 't~Hooft line can end on?  A natural candidate is the 't~Hooft-Polyakov monopole.  This suggestion, however,  faces the following difficulty.  While the field configuration near the 't~Hooft operator insertion line is singular and the operator itself is concentrated on a line, the 't~Hooft-Polyakov monopole configuration is smooth and has a definite size.  Viewed in the three-dimensional space the question is: how can a finite size smooth object screen a point like singularity?  Having obtained exact solutions, we explore this screening effect in detail in Section \ref{Sec:Analysis}.

Before we describe the field configurations we are after, we would like to emphasize that our solutions can be interpreted in a number of other theories.  They arise as half-BPS configurations in the ${\cal N}=2$ super-Yang-Mills in four dimensions. In string theory they provide a D(p+2)-brane world-volume description of a pair of parallel D(p+2)-branes connected by a finite Dp-brane with one or two semi-infinite Dp-branes ending on  the pair.  These solutions also arise in the context of a pure Yang-Mills theory at finite temperature.  In this case the field $\Phi$ in our expressions should be interpreted as a Euclidian time component $A_0$ of the gauge field and the Bogomolny equation (\ref{Eq:Bogomolny}) as a self-duality condition on ${\mathbb R}^3\times S^1$ for $S^1$  independent  configurations.

Our solutions are static and are described by the fields $\big(A(\vec{x}), \Phi(\vec{x})\big)$ on a three-dimensional space depending on the coordinate vector $\vec{x}.$  We denote the relative position of the observation point $\vec{x}$ with respect to the position of the 't~Hooft operator  by $\vec{z},$ while the relative position of the observation point with respect to the position of the center of the 't~Hooft-Polyakov monopole is denoted by $\vec{r}.$  We consider maximal symmetry breaking at infinity with the symmetry breaking scale $\lambda$ set by the Higgs field eigenvalues  at the space infinity. 

Now we spell out the exact conditions on the gauge field $A$ and the Higgs field $\Phi$ describing a single monopole in the presence of an 't~Hooft operator.  BPS monopoles \cite{Bogomolny:1975de, Prasad:1975kr} with 't~Hooft operator insertions are solutions of the Bogomolny equation \cite{Bogomolny:1975de} 
\begin{equation}
\label{Eq:Bogomolny}
F_{ij}=-\epsilon_{ijk}[D_k,\Phi],
\end{equation}
with  prescribed Dirac type singularities. Explicitly, for the minimal charge, the  condition (\ref{Eq:minimal}) implies that the Higgs field near the insertion point at $\vec{z}=\vec{0}$ is gauge equivalent to 
\begin{eqnarray}
\label{Eq:so3}
&SO(3):& \Phi_{jk}=- i \epsilon_{1jk}\frac{1}{2|\vec{z}|}+{\rm O}(|\vec{z}|^0),\\
\label{Eq:su2}
&SU(2):& \Phi_{\alpha\beta}=\sigma^3_{\alpha\beta} \frac{1}{2|\vec{z}|}+{\rm O}(|\vec{z}|^0).
\end{eqnarray}
The above conditions are written is a particular gauge for concreteness. Of course, one can perform any everywhere smooth gauge transformation to obtain an equivalent description. Here $\epsilon_{ijk}$ is a completely antisymmetric tensor and $\sigma^1, \sigma^2,$ and $\sigma^3$ are the Pauli sigma matrices.

To simplify our notation we denote the distances from the observation point to the singularity and to the monopole by $z=|\vec{z}|$ and $r=|\vec{r}|$ respectively. 
When the separation $\vec{d}=\vec{z}-\vec{r}\,$  between the 't~Hooft operator insertion and the nonabelian monopole is large, i.e. $d=|\vec{d\,} |\gg 1, 1/\lambda,$ we expect the fields $\Phi=(\Phi_{ij})$ and $A=(A_{ij})$  with
\begin{eqnarray}\label{GenSO3}
&SO(3):& \Phi_{ij}=-2i\epsilon_{ijk}\phi^k,  A_{ij}=-2i \epsilon_{ijk} A^k,\\
\label{GenSU2}
&SU(2):& \Phi=\vec{\sigma}\cdot\vec{\phi},\quad A=\vec{\sigma}\cdot\vec{A},
\end{eqnarray}
to approach those of the  't~Hooft-Polyakov BPS monopole solution \cite{'tHooft:1974qc, Polyakov:1974ek, Prasad:1975kr} 
\begin{eqnarray}\label{Eq:tHP1}
\vec{\phi} &=& \bigg(\lambda  \coth(2\lambda r) -\frac{1}{2r}\bigg)\frac{\vec{r}}{r},\\
\label{Eq:tHP2}
\vec{A} &=& \bigg(\frac{\lambda}{\sinh(2\lambda r)}-\frac{1}{2r}\bigg)\frac{\vec{r}\times d\vec{x}}{r}.
\end{eqnarray}

We present the solutions satisfying the above conditions in Section \ref{Sec:Solutions} and analyze them in Section \ref{Sec:Analysis}.
We used the technique of the Nahm transform \cite{Nahm:1982jb}, outlined in the Appendix, to obtain the explicit solutions presented here.  More general solutions with two singularities appear in \cite{ChD}.  The solutions below are exact and have been explicitly verified analytically and numerically.

\section{Solutions}\label{Sec:Solutions}
It is convenient to introduce  ${\cal D}=2zd+2\vec{z}\cdot\vec{d}=(z+d)^2-r^2,$ and 
to use the vector-valued functions $\vec{\phi}=(\phi^1,\phi^2,\phi^3)$ and $\vec{A}=(A^1,A^2,A^3).$  Then the monopole solutions of the Bogomolny Eq.~(\ref{Eq:Bogomolny}) satisfying the boundary conditions (\ref{Eq:so3}) and (\ref{Eq:su2}) are provided by Eq.(\ref{GenSO3}) and Eq.(\ref{GenSU2}) above with $\vec{\phi}$ and $\vec{A}$ given respectively as follows:

{$\mathbf{SO(3)}$ \bf  Case:}
\begin{eqnarray}\nonumber
\vec{\phi}&=&\left(\bigg(\lambda+\frac{1}{4 z}\bigg)\frac{k}{l}-\frac{1}{2 r}\right)\frac{\vec{r}}{r}-\frac{r}{2 z l \sqrt{{\cal D}}}\left(\vec{d}-\frac{\vec{r}\cdot\vec{d}}{r^2}\vec{r}\right),\\
\label{Eq:SO(3)fields}
\vec{A}&=&\left(\bigg(\lambda+\frac{z+d}{2{\cal D}} \bigg)\frac{\sqrt{{\cal D}}}{l}-\frac{1}{2r}\right)\frac{\vec{r}\times d\vec{x}}{r}\\
&&-\frac{r}{2 l \sqrt{{\cal D}}}\left(\frac{\vec{z}\times d\vec{x}}{z}+ \bigg(\frac{k}{\sqrt{\cal D}}-1 \bigg) \frac{(\vec{r}\cdot(\vec{z}\times d\vec{x}))}{r z}\frac{\vec{r}}{r}\right),\nonumber
\end{eqnarray}
where
\begin{eqnarray}
l &=& (z+d) \sinh(2\lambda r)+r\cosh(2\lambda r),\\
k &=& (z+d) \cosh(2\lambda r)+r\sinh(2\lambda r).
\end{eqnarray}

{$\mathbf{SU(2)}$ \bf Case:}
\begin{eqnarray}\nonumber
\vec{\phi}&=&\left(\bigg(\lambda+\frac{1}{2z} \bigg)\frac{\cal K}{{\cal L}}-\frac{1}{2r}\right)\frac{\vec{r}}{r}-\frac{r}{z{\cal L}}\left(\vec{d}-\frac{\vec{r}\cdot\vec{d}}{r^2}\vec{r}\right),\\
\label{Eq:SU(2)fields}
\vec{A} &=&\left(\bigg(\lambda+\frac{z+d}{{\cal D}} \bigg)\frac{{\cal D}}{{\cal L} }-\frac{1}{2r}\right)\frac{\vec{r}\times d\vec{x}}{r}\\
\nonumber
&&-\frac{r}{{\cal L}}\left(\frac{\vec{z}\times d\vec{x}}{z}+ \bigg(\frac{{\cal K}}{{\cal D}}-1 \bigg)\frac{(\vec{r}\cdot(\vec{z}\times d\vec{x}))}{rz}\frac{\vec{r}}{r}\right),
\end{eqnarray}
where
\begin{eqnarray}
{\cal L} &=& ((z+d)^2+r^2) \sinh(2\lambda r)+2r(z+d)\cosh(2\lambda r),\\
{\cal K} &=& ((z+d)^2+r^2) \cosh(2\lambda r)+2r(z+d)\sinh(2\lambda r).
\end{eqnarray}

\section{Analysis}\label{Sec:Analysis}
The form of the expressions (\ref{Eq:SO(3)fields}) and (\ref{Eq:SU(2)fields}) makes the large separation limit  transparent. Indeed, in this limit the fields near the 't~Hooft operator insertion $(d\rightarrow\infty, z\   {\rm finite})$ reproduce those of Eqs.~(\ref{Eq:so3},\ref{Eq:su2}), while near the monopole core $(d\rightarrow\infty, r\ {\rm finite})$ they approach the 't~Hooft-Polyakov solution (\ref{Eq:tHP1}, \ref{Eq:tHP2}).

There is a substantial difference in the behavior of these solutions as we decrease $d$ and the nonabelian monopole and the 't~Hooft operator collide. The $SO(3)$ solution at $d=0$ becomes another 't~Hooft operator with the Higgs field 
\begin {equation}\label{Eq:final}
\vec{\phi}=\left(\lambda-\frac{1}{4r}\right)\frac{\vec{r}}{r},
\end{equation}
while the $SU(2)$ solution in this limit becomes trivial  with $F=0,$ and  $\Phi$  constant.  The latter illustrates the screening effect, in which a nonabelian monopole completely screens the point-like singularity of the 't~Hooft operator.  
A priori one might think such screening impossible since the 't~Hooft-Polyakov monopole has a finite size of order $1/\lambda$ and finite energy density in the core, while the Dirac singularity of the 't~Hooft operator is pointlike with the energy density divergent at one point. 
Electric-magnetic duality, however, suggests such screening as a possible dual explanation of screening of Wilson line operators by quarks. Our solution (\ref{Eq:SU(2)fields}) resolves this seeming contradiction as we now explain.    

\begin{figure*}[h]
\subfigure[\ Gauge group $SO(3).$]{
\includegraphics[width=60mm]{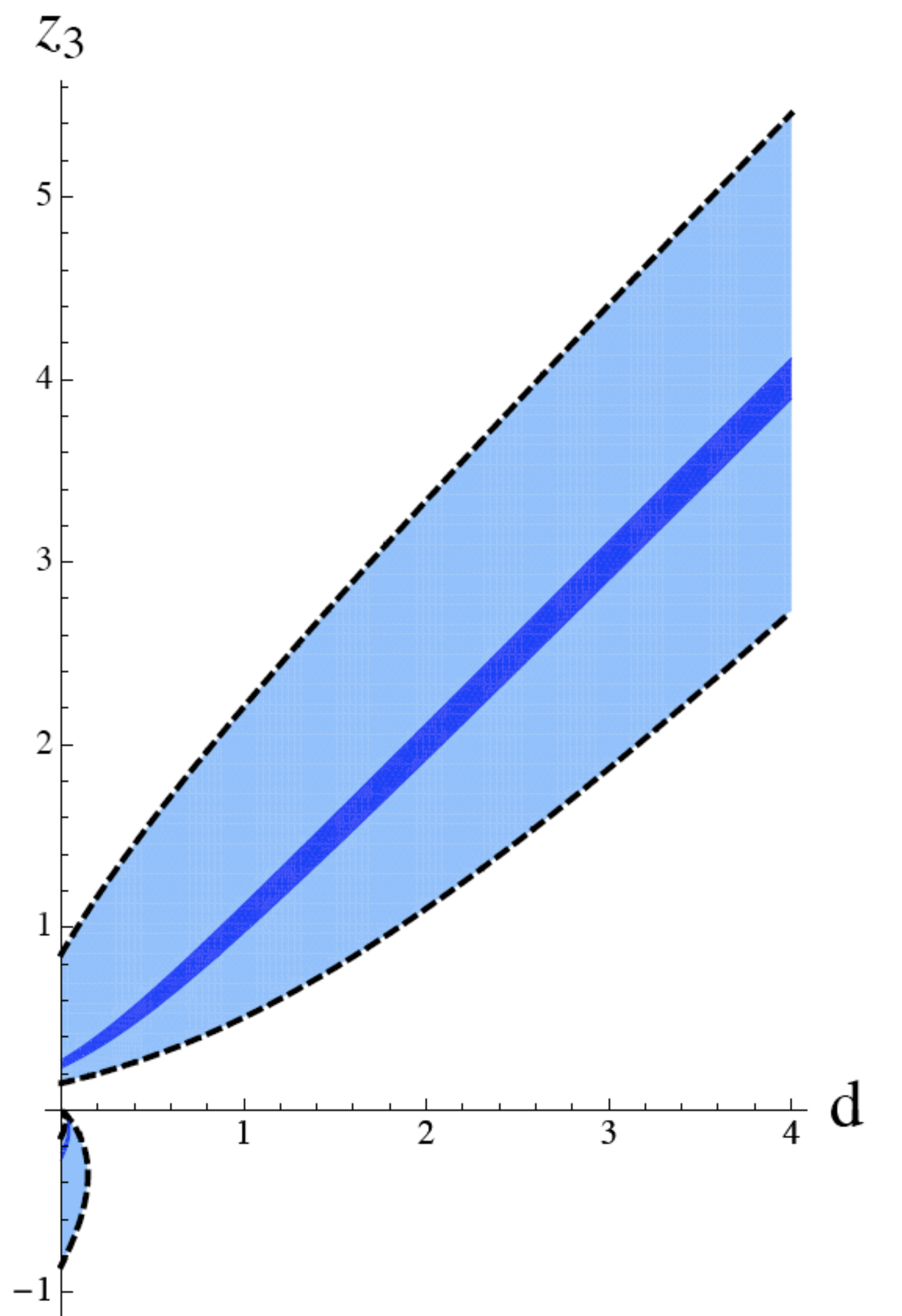}}
\hfill
\subfigure[\ Gauge group $SU(2).$]{
\includegraphics[width=60mm]{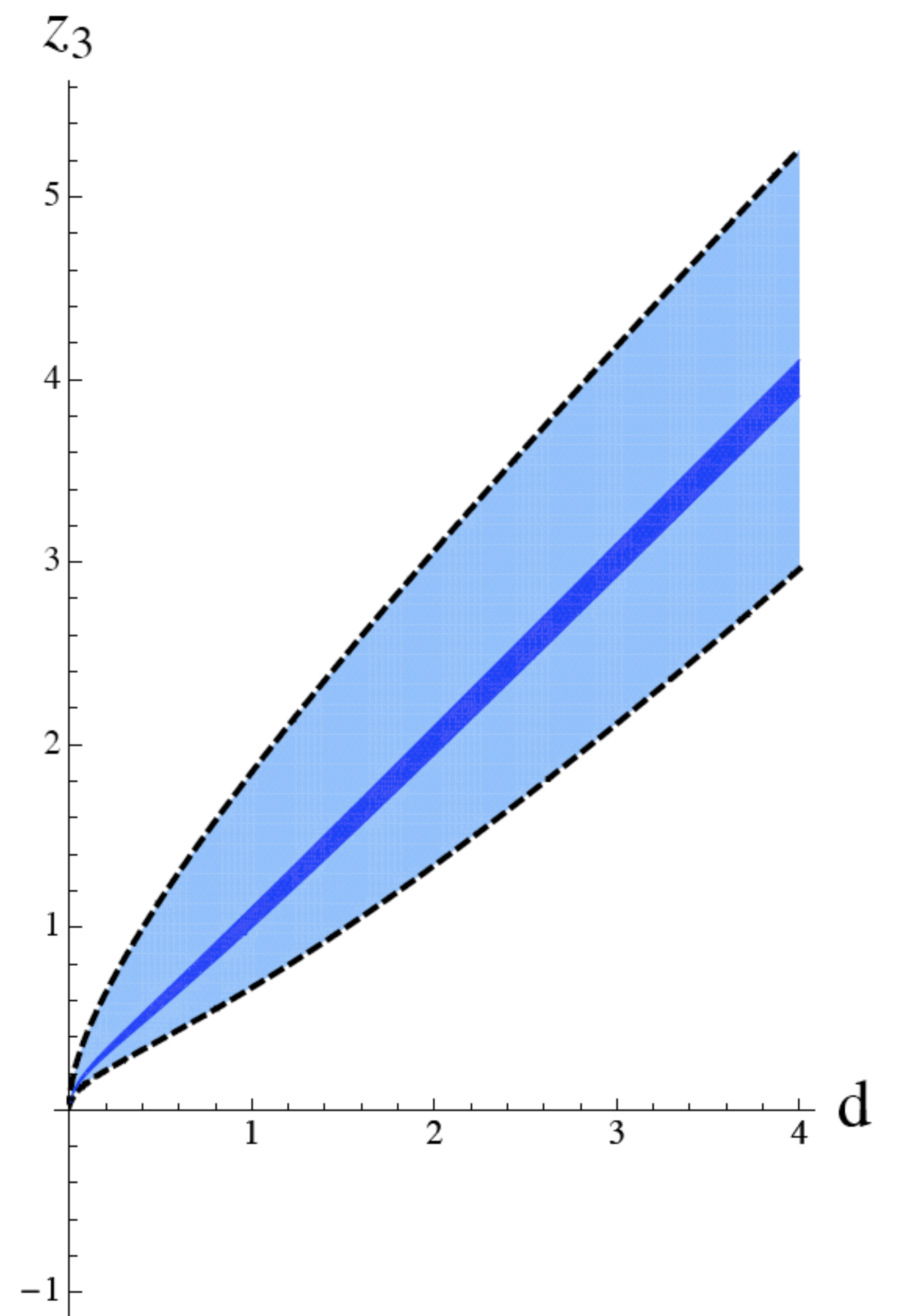}}
\caption{\label{Contour} Higgs field profiles for  $\lambda=1.$ Dashed lines correspond to $|\Phi|^2=\frac{1}{2},$ the shaded area $|\Phi|^2\leq\frac{1}{2},$ and the dark region indicates the position of the monopole core.}
\end{figure*}
There is a number of ways to explore the size of the monopole. One is via the energy density distribution ${\cal E}\sim\frac{1}{2}{\rm Tr}\Big(F^2+(D\Phi)^2\Big)=\Big(\partial^2_{x_1}+\partial^2_{x_2}+\partial^2_{x_3}\Big){\rm Tr} \Phi^2$ and another is by how much the gauge symmetry is broken by the Higgs field. In particular, we might think of the position of the monopole as the point where the Higgs field vanishes and the gauge symmetry is fully restored.  A word of caution is due here.  Even though the parameter $\vec{d}$ is a good indication of the monopole relative position when  $d$ is large, it is not the point where the Higgs field vanishes, rather, at $\vec{d}$ (i.e. at $\vec{r}=\vec{0}$) we have
\begin{equation}
SO(3): |\vec{\phi}|=\frac{1}{4d(1+4\lambda d)},\ SU(2): |\vec{\phi}|=\frac{1}{4d(1+2\lambda d)}.
\end{equation}
For the two gauge groups the profiles of $|\Phi|^2=\frac{1}{2}{\rm Tr}\Phi^2$  at large separation parameter $d$ look remarkably similar to each other.  They do differ drastically, however, for small values of $d.$  One can infer from Figs.~\ref{Contour}(a) and \ref{Contour}(b) how the position and size of the monopole vary with $d$.   The shaded areas in these graphs corresponds to the values of $|\Phi|^2\leq 1/2$ and we choose the asymptotic condition to be $|\Phi(\infty)|=\lambda=1.$  The coordinate $z_3$ is chosen along the line originating at the 't~Hooft operator and passing through the monopole. The dark area in the middle corresponds to the values of $|\Phi|^2<0.007,$ giving a good indication of the position of the monopole center. 

In the case of the $SU(2)$ gauge group, Fig.~\ref{Contour}(b), the monopole shrinks to zero size as it approaches the singularity and screens the 't~Hooft operator completely.  In the case of the $SO(3)$ gauge group, as the parameter $d\rightarrow 0,$ the longitudinal size of the monopole decreases somewhat, approaching a constant. Instead of the monopole shrinking to zero size, we observe it spreading transversely and eventually encircling the singularity.  This is also indicated by the expression (\ref{Eq:final}) representing the limiting configuration to be an 't~Hooft operator surrounded by a sphere of vanishing Higgs field at $z=\frac{1}{2\lambda}.$

\begin{figure*}[h!]
\subfigure[\ Gauge group $SO(3).$]{
\includegraphics[width=60mm]{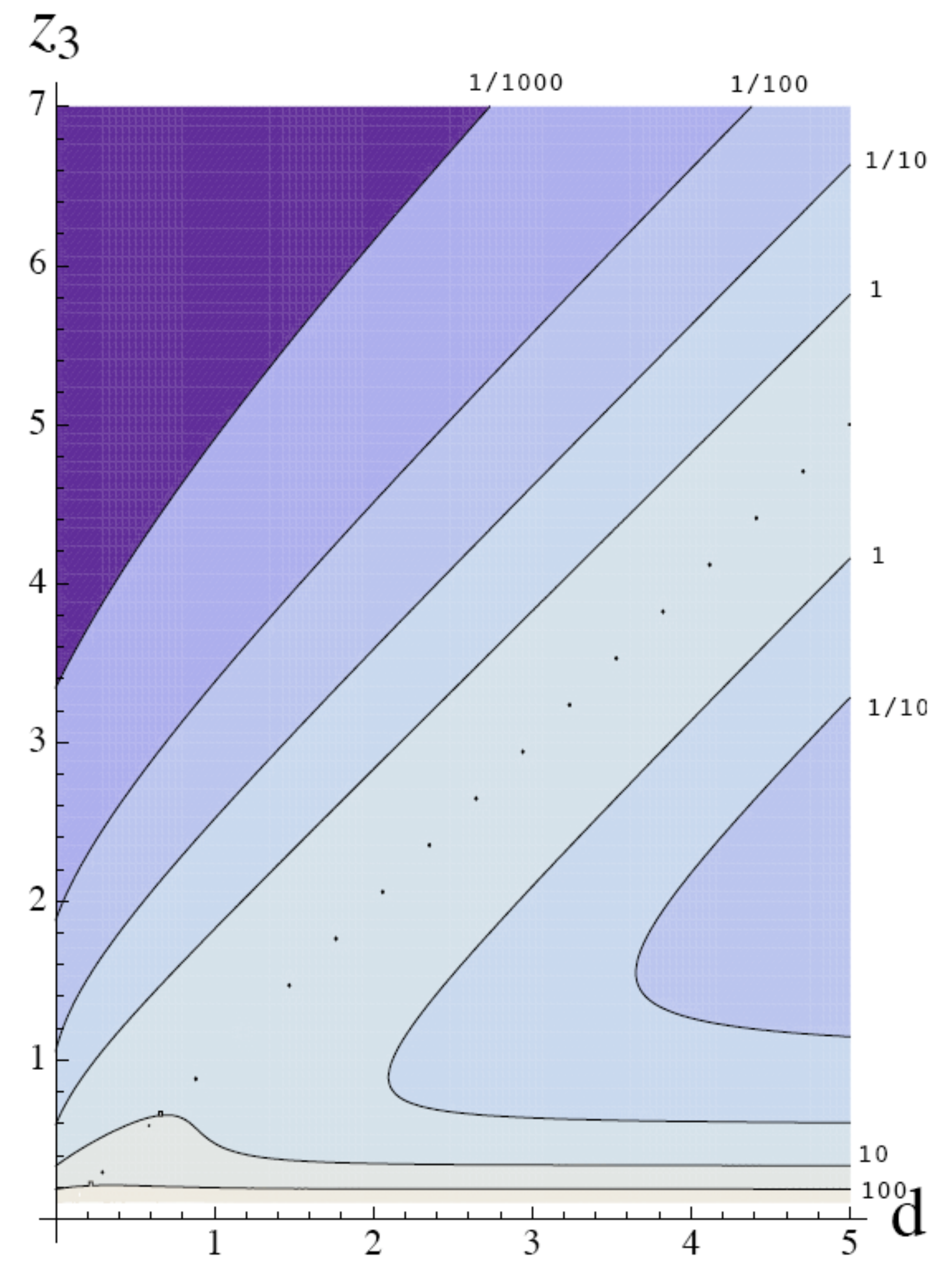}}
\hfill
\subfigure[\ Gauge group $SU(2).$]{
\includegraphics[width=60mm]{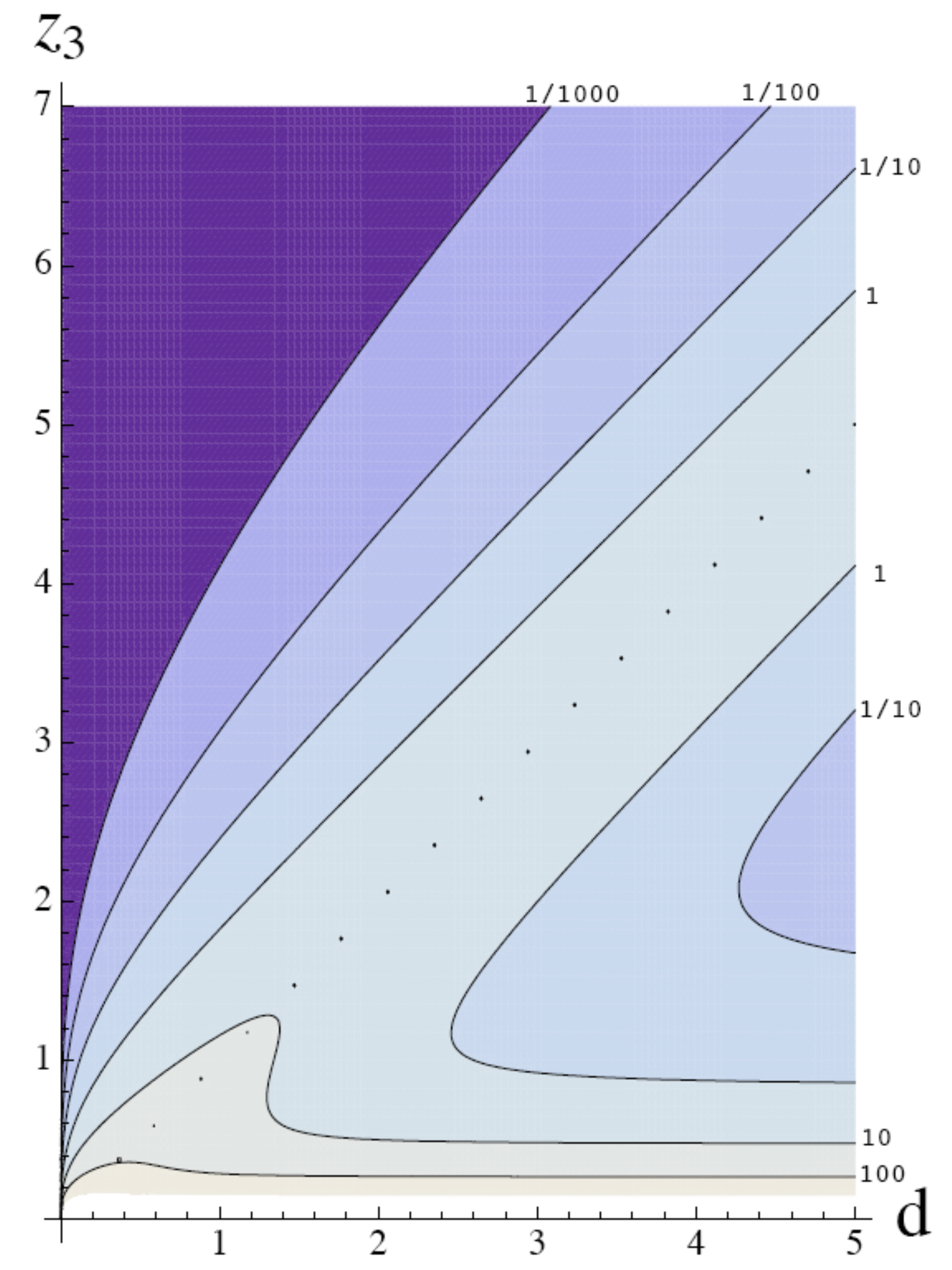}}
\caption{\label{EnergyDensity} Energy density contour plots for $  \lambda=1.$}
\end{figure*}

Energy density plots of Figs.~\ref{EnergyDensity}(a) and \ref{EnergyDensity}(b) support this picture.  They present the contour levels of $\Delta \Big.\vec{\phi}\  \Big.^2,$ which is proportional to the energy density ${\cal E},$ due to Bogomolny Eq.~(\ref{Eq:Bogomolny}) and the Bianchi identity.  As $d\rightarrow 0$ the energy density for the $SO(3)$ gauge group case approaches a steady distribution diverging at the origin, while for the $SU(2)$ case the energy density decreases uniformly and, at $d=0,$ vanishes everywhere.

\section{Comments}
One can view the two solutions discussed here as part of the same picture since every $SU(2)$ solution can be viewed as one with the gauge group $SO(3)$ by factoring out the center of the group. Thus, we can reinterpret Eq.(\ref{Eq:SU(2)fields}) combined with Eq.(\ref{GenSO3}) as an $SO(3)$ monopole with the charge two 't~Hooft operator insertion.  Such an interpretation provides a topological reason for our observations in the previous section.  For the gauge group $SO(3)$ the 't~Hooft charge takes value in ${\mathbb Z}/2{\mathbb Z}.$  The solution (\ref{GenSO3}, \ref{Eq:SO(3)fields}) has 't~Hooft charge one operator insertion and thus it is topologically protected. The solution (\ref{GenSO3}, \ref{Eq:SU(2)fields}), however, has Dirac charge two and vanishing 't Hooft charge operator insertion which, as a result, can be screened completely.  One can view the latter charge two configuration as a limit of two minimal 't~Hooft operators approaching each other.  We explore such a limit in more detail in  \cite{ChD}.

\section*{Acknowledgments}
BD is supported by the Irish Research Council for Science Engineering and Technology: funded by the National Development Plan. The work of SCh is supported by the Science Foundation Ireland Grant No. 06/RFP/MAT050 and by the European Commision FP6 program MRTN-CT-2004-005104.

\section*{Appendix}
Here we briefly outline the Nahm transform techniques, which led us to the solutions presented. For the configuration of one nonabelian $SU(2)$ monopole positioned at $\vec{T}_{'tHP}$ in the presence of a minimally charged 't~Hooft operator at $\vec{T}_D,$ we denote the relative positions by $\vec{r}=\vec{x}-\vec{T}_{'tHP},\vec{z}=\vec{x}-\vec{T}_{D},\vec{d}=\vec{T}_{'tHP}-\vec{T}_{D}.$ We also let $r=|\vec{r}|,\ z=|\vec{z}|,$ and $d=|\vec{d}|.$

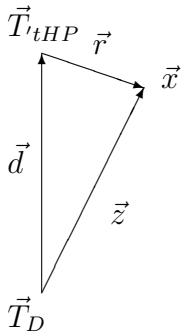
\begin{figure}[h!]
\label{3pointdiagram}
\setlength{\unitlength}{1.1em}
\begin{center}
\begin{picture}(10,10)(0,0)
\put(5,2){\vector(0,1){7}}
\put(5,9){\vector(3,-1){3}}
\put(5,2){\vector(1,2){3}}
\put(4,5.5){$\vec{d}$}
\put(6.5,9){$\vec{r}$}
\put(7,4){$\vec{z}$}
\put(4,1){$\vec{T}_{D}$}
\put(4,9.5){$\vec{T}_{'tHP}$}
\put(8.5,8){$\vec{x}$}
\end{picture}
\end{center}
\caption{Positions of the monopole $\vec{T}_{'tHP}$, the 't~Hooft operator $\vec{T}_{D}$, and the observation point $\vec{x}.$} 
\end{figure}

\begin{figure}[h]
\label{BraneDiagram}
\setlength{\unitlength}{1.1em}
\begin{center}
\begin{picture}(17,7)(0,0)
\put(2,1){\line(0,1){6}}
\put(1,0){$-\lambda$}
\put(2,5){\line(1,0){5}}
\put(7,1){\line(0,1){6}}
\put(7,0){$\lambda$}
\put(7,4){\line(1,0){8}}
\put(11,4.3){$\vec{T}_{D}$}
\put(4,5.3){$\vec{T}_{'tHP}$}
\end{picture}
\end{center}
\caption{Brane diagram signifying the Nahm data on an interval $(-\lambda, \lambda)$ and a semi-infinite interval $(\lambda, \infty).$ This diagram depicts the Nahm data defining an $SU(2)$ monopole in the presence of one minimal charge 't~Hooft operator.}
\end{figure}
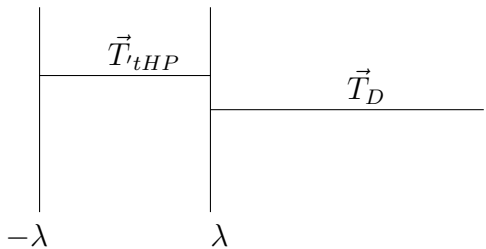
The relevant Nahm data for this case is given by a piecewise constant vector-valued function
\begin{equation}
\label{1 sing data}
\vec{T}(s) = \left\{
\begin{array}{lcl}
\vec{T}_{'t HP}\in{\mathbb R}^3 & \mbox{for} & s\in(-\lambda,\lambda)\\
\vec{T}_{D}\in{\mathbb R}^3 & \mbox{for} & s>\lambda
\end{array}
\right.,
\end{equation}
and a 2-component spinor $f_+$ satisfying
$\frac{1}{2}f_+^\dagger\vec{\sigma} f_+=\vec{T}_{'tHP}-\vec{T}_{D}=\vec{d}.$ It follows that it satisfies $f_+ f_+^\dagger=d+\sdv{d}.$  For convenience we also define spinors $\zeta_+$ and $\zeta_-$ satisfying $\zeta_\pm\zeta_\pm^\dagger=z+\sdv{z}.$

The next step of the Nahm transform is for any observation point $\vec{x}$ to define the following Weyl operator
\begin{equation}
\Dslash_x^\dagger\Psi = \bigg({\mathbb I}_{2\times 2}\otimes \frac{d}{ds}+\sum_{j=1}^3 \sigma_j\otimes (T_j(s)-x^j)\bigg)\psi(s)-\delta(s-\lambda)f_+\Delta_\lambda,
\end{equation}
acting on  $\Psi=\left(\begin{array}{c}\psi(s)\\ \Delta_\lambda\end{array}\right),$ 
and to find solutions of the equation $\Dslash_x^\dagger\Psi_n=0.$
In our case it has two dimensional space of solutions.  We organize an orthonormal basis $(\Psi_1, \Psi_2)$ in this space into a matrix $\Psi=(\Psi_1, \Psi_2)$ so that the $n$-th column is the $n$-th element of this basis.

An explicit solution $\Psi=\left(\begin{array}{c}\psi(s)\\ \Delta_\lambda\end{array}\right)$ we find is
\begin{eqnarray}
\psi(s) &=& \left\{
\begin{array}{lcl}
\sqrt{\frac{r}{\sh}}e^{\sdv{r} s} N & \mbox{for} & s\in[-\lambda,\lambda)\\
e^{\sdv{z}(s-\lambda)}\frac{\zeta_-f_-^\dagger}{f_-^\dagger\zeta_-}\sqrt{\frac{r}{\sh}}e^{\sdv{r} \lambda}N & \mbox{for} & s>\lambda
\end{array}
\right.,\\
\Delta_\lambda&=&-\frac{\zeta_+^\dagger}{\zeta_+^\dagger f_+}\sqrt{\frac{r}{\sh}}e^{\sdv{r} \lambda}N,
\end{eqnarray}
where $N$ is the normalization constant given by
\begin{eqnarray}
N &=& \frac{\sqrt{z+d+\sqrt{{\cal D}}}-\sqrt{z+d-\sqrt{{\cal D}}}\sdv{r}/r}{\sqrt{2(z+d+r\coth2\lambda r)}},\\
\label{D}
{\cal D} &=& (z+d)^2 - r^2 = 2 z d + 2 \vec{z}\cdot\vec{d},
\end{eqnarray}
chosen to ensure that that the solutions are ortho-normalized
\begin{equation}
(\Psi^\dagger, \Psi)= \Delta_\lambda^\dagger\Delta_\lambda+\int_{-\lambda}^{\infty} ds \psi^\dagger(s)\psi(s)={\mathbb I_{2\times 2}}.
\end{equation}

The Higgs field and connection are then given by
\begin{eqnarray}
\Phi &=& \lambda\Delta^\dagger_\lambda\Delta_\lambda + \int_{-\lambda}^\infty ds \psi^\dagger s\psi,\\
A &=& i\vec{dx}\cdot\bigg( \Delta^\dagger_\lambda\vec{\nabla}_x\Delta_\lambda+ \int_{-\lambda}^\infty ds \psi^\dagger \vec{\nabla}_x \psi \bigg),
\end{eqnarray}
giving directly our solution Eqs.~(\ref{GenSU2}, \ref{Eq:SU(2)fields}).

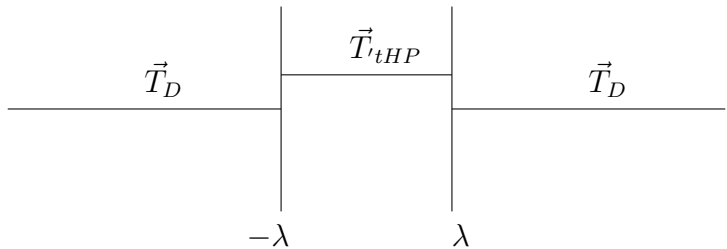
\begin{figure}

\begin{center}
\setlength{\unitlength}{1.1em}
\begin{picture}(17,7)(0,0)

\put(5,1){\line(0,1){6}}
\put(4,0){$-\lambda$}
\put(5,5){\line(1,0){5}}
\put(5,4){\line(-1,0){8}}
\put(10,1){\line(0,1){6}}
\put(10,0){$\lambda$}
\put(10,4){\line(1,0){8}}
\put(7,5.5){$\vec{T}_{'tHP}$}
\put(1,4.5){$\vec{T}_{D}$}
\put(14,4.5){$\vec{T}_{D}$}

\end{picture}
\caption{Brane diagram for the Nahm data corresponding to an $SO(3)$ monopole at $\vec{T}_{'tHP}$ and one minimal charge 't~Hooft operator at $\vec{T}_D.$ The data is defined on ${\mathbb R}$ and is continuous outside the points $s=-\lambda,\ \lambda.$}
\label{Fig:Both}
\end{center}
\end{figure}

Formulating the Nahm transform for the $SO(3)$ case is completely analogous, even though the computations are a little more tedious. In this case the Nahm data is defined on the whole real line with discontinuities at $s=\pm\lambda.$ Also,  besides the $f_+,$ spinor accounting for the discontinuity in the Nahm data at $s=\lambda,$  we also have a spinor $f_-$ corresponding to the discontinuity at $s=-\lambda.$ We present the brane configuration for the corresponding Nahm data in Fig. \ref{Fig:Both}. The Weyl operator acts on $\Psi=\left(\begin{array}{c}\psi(s)\\ \Delta_\lambda\\ \Delta_{-\lambda}\end{array}\right)$ and it has a term containing $\delta(s+\lambda) f_-.$  An ortho-normalized solution of $\Dslash_x^\dagger\Psi$ leads to the expressions  in Eqs. (\ref{GenSO3}, \ref{Eq:SO(3)fields}).

\end{document}